\documentclass[twocolumn]{aastex631}

\usepackage{amsmath}

\newcommand{\tet}{$\Theta^2$A\,Ori}
\newcommand{\hef}{$^{4}$He}
\newcommand{\het}{$^{3}$He}
\newcommand{\heiso}{\het/\hef}

\newcommand{\HeIs}{He\,\textsc{i}*}
\newcommand{\HetIs}{$^{3}$He\,\textsc{i}*}
\newcommand{\HefIs}{$^{4}$He\,\textsc{i}*}

\newcommand{\HeII}{He\,\textsc{ii}}
\newcommand{\HI}{H\,\textsc{i}}
\newcommand{\HII}{H\,\textsc{ii}}
\newcommand{\obhh}{$\Omega_{\rm B,0}\,h^{2}$}

\newcommand{\vice}{\texttt{VICE}}
\newcommand{\mathsc}[1]{{\normalfont\textsc{#1}}}

\usepackage{xcolor}

\shorttitle{The primordial helium isotope ratio}
\shortauthors{Cooke et al.}

\graphicspath{{./}{figures/}}

\begin{document}

\title{Towards a measurement of the primordial helium isotope ratio\footnote{Based on observations collected at the European Organisation for Astronomical Research in the Southern Hemisphere under ESO programme(s) 081.D-0854(B), 091.C-0851(A), 107.22U1.001, 110.23QS.001, 110.23QS.002, 112.25D8.001, 112.25D8.002, 194.C-0833(A), 194.C-0833(D).}}

\author[0000-0001-7653-5827]{Ryan J. Cooke}
\correspondingauthor{Ryan J. Cooke}
\email{ryan.j.cooke@durham.ac.uk}
\affiliation{Centre for Extragalactic Astronomy, Durham University, Durham DH1 3LE, UK}
\affiliation{Department of Physics, Durham University, South Road, Durham DH1 3LE, UK}

\author[0000-0002-6534-8783]{James W. Johnson}
\affiliation{Carnegie Science Observatories, 813 Santa Barbara Street, Pasadena, CA 91101, USA}

\author[0000-0002-5777-1629]{Pasquier Noterdaeme}
\affiliation{Institut d'Astrophysique de Paris, UMR 7095, CNRS and SU, 98bis bd Arago, 75014 Paris, France}

\author[0000-0002-5139-4359]{Max Pettini}
\affiliation{Institute of Astronomy, University of Cambridge, Madingley Road, Cambridge, CB3 0HA, UK}

\author[0000-0003-3174-7054]{Louise Welsh}
\affiliation{Centre for Extragalactic Astronomy, Durham University, Durham DH1 3LE, UK}
\affiliation{Department of Physics, Durham University, South Road, Durham DH1 3LE, UK}

\author[0009-0008-2352-8314]{Aldric Wong}
\affiliation{Centre for Extragalactic Astronomy, Durham University, Durham DH1 3LE, UK}
\affiliation{Department of Physics, Durham University, South Road, Durham DH1 3LE, UK}

\author[0000-0002-4288-599X]{Celine Peroux}
\affiliation{European Southern Observatory, Karl-Schwarzschildstrasse 2, D-85748 Garching bei M\"unchen, Germany}
\affiliation{Aix Marseille Universit\'e, CNRS, LAM (Laboratoire d’Astrophysique de Marseille) UMR 7326, F-13388 Marseille, France}

\begin{abstract}
We report the discovery of two metastable neutral helium (\HeIs) absorbers in the Milky Way, and use the upgraded CRyogenic InfraRed Echelle Spectrograph on the Very Large Telescope to determine the helium isotope ratio, \heiso, along these sightlines. We have also obtained deeper observations of a third sightline to report a $\lesssim4\%$ precision measure of \heiso\ in the Orion Nebula. These data have allowed us to place a $2\sigma$ limit on the time-variability of \HeIs\ absorption in the Orion nebula, ${\rm d}\log_{10} [N({\rm He\,\textsc{i}}^{*})/{\rm cm}^{-2}]/{\rm d}t\leq7.2\times10^{-4}~{\rm dex~yr}^{-1}$ ($<0.17\%~{\rm yr}^{-1}$), suggesting that these absorbers are in radiative equilibrium. We compute new galactic chemical evolution models of the Milky Way, and use our observations to infer the primordial helium isotope ratio and a scaling factor for the yields reported by nucleosynthesis calculations. Based on the data and models that we report here, we infer a best-fit value (\heiso)$_{\rm P}=(1.15^{+0.24}_{-0.21})\times10^{-4}$, which agrees with Big Bang nucleosynthesis calculations that assume the Standard Model of particle physics in combination with the baryon density inferred from the cosmic microwave background temperature fluctuations. We infer the stellar yield scale relative to the solar metallicity, $y/Z_{\odot}=2.12^{+0.31}_{-0.29}$, which is somewhat higher than previously found. Finally, we note that the forthcoming extremely large telescopes are poised to determine \heiso\ in more metal-poor environments, to secure a model-independent determination of the primordial value.
\end{abstract}


\section{Introduction}
\label{sec:intro}

Within just six minutes after the Big Bang, $99.9\%$ of all primordial helium-4 (\hef) in the Universe was forged. In contrast, it took almost 10 hours before the lighter stable isotope of helium (\het) reached its freeze-out value. The freeze-out values of these primordially-produced nuclides depend primarily on the baryon-to-photon ratio, which is directly proportional to the baryon density of the Universe \citep{Steigman2006}. This brief period of nucleosynthesis, generally referred to as primordial nucleosynthesis or Big Bang nucleosynthesis (BBN), is only capable of producing the lightest nuclides in measurable abundances (for a recent review, see \citealt{Cooke2026}). In particular, there has been a great effort to measure
the primordial abundance of deuterium \citep[D/H;][]{Cooke18,Guarneri2024,Kislitsyn2024},
\hef/H \citep[$y_{\rm P}$;][]{Hsyu20,Kurichin2021,Yanagisawa2025,Aver2026}, and
lithium-7 \citep[$^{7}$Li/H; see][for an observational overview]{Norris2023}.\footnote{Note that some tritium and beryllium-7 are also made, but these decay to helium-3 and lithium-7 $\sim12$ years and $\sim600$ years after the Big Bang, respectively \citep{KhatriSunyaev2011}.}

Unlike these more commonly observed nuclides, there are relatively fewer attempts to measure the primordial abundance of \het/H. For the most part, this is thanks to the difficulty of obtaining reliable detections of \het. In almost all environments, the heavier isotope, \hef, is about 10,000 times more abundant than \het. Furthermore, the isotope shift is generally $\lesssim15~{\rm km~s}^{-1}$ \citep{MorWuDra06}. One way to overcome these difficulties is to observe the 8.7 GHz spin-flip transition of \het$^{+}$. This hyperfine structure transition arises because \het\ has a non-zero nuclear magnetic moment. Meanwhile, \hef\ has a zero nuclear magnetic moment, and therefore does not exhibit a spin-flip transition.

To date, the \het$^{+}$ 8.7~GHz transition has been used to measure the \het/H abundance of five Milky Way \HII\ regions \citep{BalBan18}.\footnote{There are also a few potential detections of \het$^{+}$ in planetary nebulae, but these have not been confirmed with higher sensitivity observations \citep{BanBal21,Balser2022}.} One of these \HII\ regions, dubbed `Sh2-209', has a well-characterized simple structure and resides in the outskirts of the Milky Way, with a galactocentric radius $R_{\rm gal}\simeq12.7~{\rm kpc}$.\footnote{Before the \textit{Gaia} satellite measured the distance to the resident stars of Sh2-209, this \HII\ region was thought to be at $R_{\rm gal}=16~{\rm kpc}$.} In the outskirts of the Milky Way, galactic chemical evolution models predict that the \het\ abundance is within $\sim20$ percent of the primordial value \citep{Romano2003,Lagarde2012,Cooke2022,WellerWeinberg2026}. \citet{BalBan18} determine that this \HII\ region exhibits a \het/H abundance that is consistent with `Standard Model' BBN calculations (i.e. assuming the Standard cosmological model and the Standard Model of particle physics, in combination with the baryon density measured from the Cosmic Microwave Background temperature fluctuations; hereafter, referred to as the `Standard Model value').

The same spin-flip transition may also be detectable in the near-primordial regions of the intergalactic medium along the line-of-sight to radio loud quasars during \HeII\ reionization \citep{McQSwi2009,Takeuchi14,Khullar20}. These intergalactic clouds of mostly ionized gas are expected to imprint a forest of redshifted \het$^{+}$ 8.7~GHz absorption lines on the spectrum of the background quasar. While this technique has potential given the near-pristine composition of the gas, the absorption is expected to be extremely weak, so current facilities are not sensitive enough to attempt this experiment. Furthermore, this approach would require an ionization correction to convert the abundance of $^{3}$\HeII/\HI\ to the abundance of \het/H.

\het\ has also been detected in several environments of the Solar System, where its abundance is usually measured relative to \hef. The most reliable determination of the proto-solar helium isotope ratio, \het/\hef, is based on a measurement of Jupiter's atmospheric composition using the \emph{Galileo Probe Mass Spectrometer} ($1.66\pm0.05\times10^{-4}$; \citealt{Mah1998}). \het/\hef\ has also been measured in Earth's mantle \citep{Peron18}, meteorites \citep{Buse00,Kri21}, and solar wind particles based on the \emph{Genesis} mission \citep{Heber2012}. Several \het-rich solar energetic particle events have also been detected with the \emph{Parker Solar Probe} \citep{Wiedenbeck2020,Hart2024}. Beyond the Solar System, the present day \het/\hef\ value of the Local Interstellar Cloud has also been measured ($1.62\pm0.29\times10^{-4}$; \citealt{Buse06}) using foils that were exposed to the interstellar neutral particle flux.

Recently, a new technique was used to measure the helium isotope ratio of the Orion Nebula \citep{Cooke2022}. This approach uses quiescent metastable neutral helium lines, \HeIs, which are seen in absorption against the light of a bright O star. The absorption is likely due to a dense cloud of gas along the line-of-sight that is being illuminated by the strong radiation field of the nearby hot stars. These authors measured a value $^{3}{\rm He}/^{4}{\rm He} = (1.77 \pm 0.13) \times 10^{-4}$, which is just $\sim40\%$ above the Standard Model primordial helium isotope ratio,\footnote{The quoted `Standard Model' value was calculated using the \texttt{LINX} code \citep{Giovanetti2025} combined with the \texttt{PArthENoPE} rates \citep{Gariazzo2022}, assuming the \citet{Planck2018} value of the baryon density, $100\,\Omega_{\rm B,0}\,h^{2}=2.233\pm0.015$. This Standard Model value of (\heiso)$_{\rm P}$ is in good agreement with other BBN calculations that adopt different choices for the nuclear reaction rates and BBN input physics. For example, ($^{3}{\rm He}/^{4}{\rm He})_{\rm P} = (1.267\pm0.017)\times10^{-4}$ using \texttt{PRIMAT} \citep{Pitrou21}.} $(^{3}{\rm He}/^{4}{\rm He})_{\rm P} = (1.262\pm0.017)\times10^{-4}$. This suggests that there is relatively little \het\ produced by stars, even in a galaxy that is as metal-rich as the Milky Way.

Such metastable helium absorption lines have been known for almost a century \citep{Wilson37}. However, despite this early discovery, there are only a small handful of \HeIs\ absorbers currently known. The full complement of known systems includes five sightlines towards Orion \citep{Odell93,Oud97}, a few sightlines towards NGC\,6611 \citep{Evans05}, and towards $\zeta$\,Oph \citep{GalKre12}. Extragalactic \HeIs\ absorption has also been detected along the sightline to a gamma ray burst \citep{Fynbo14}, and many broad absorption line quasars \citep{Liu2015}. Unfortunately, gamma ray bursts are transient, and it is difficult to secure high spectral resolution data ($R\sim100,000$) with sufficient signal-to-noise ratio to detect metastable \het\ absorption. Meanwhile, broad absorption line quasars are relatively faint given the typically high S/N and spectral resolution that are required to detect \HetIs. Furthermore, broad absorption line quasars show complex kinematics; their absorption lines often exhibit a lot of structure that would complicate a measurement of \het.

For the foreseeable future, the best opportunity to infer the primordial helium isotope ratio is to understand the production of \het\ in the Milky Way environment. Towards this goal, we have undertaken a dedicated study to identify new \HeIs\ absorbers and deduce the helium isotope ratio in a range of Galactic environments. Measurements of non-metal isotopes also provide unique information that can break the degeneracy between the scale of the stellar yields reported by nucleosynthesis calculations and the rate with which gas is ejected from the interstellar medium 
\citep{Hartwick1976, Johnson2023dwarfs, Sandford2024}.
Ideally, the scale of the stellar yields should be a value of 1, which would imply that nucleosynthesis calculations accurately determine the true yield. However, metal yields from stellar populations are uncertain by factors of $\sim2-3$ due to uncertainties associated with, e.g., convection \citep{Chieffi2001, Ventura2013, Issa2025}, rotational mixing \citep{Frischknecht2016, Limongi2018}, mass loss rates \citep{Sukhbold2016, Beasor2020}, nuclear reaction rates \citep{Herwig2004, Herwig2005}, binarity \citep{Farmer2021}, and failed supernovae \citep{Ertl2016, Griffith2021, Bruenn2023, Vartanyan2023}.
Increases in metal yields can be compensated by proportional increases in the rate at which metals are lost in galactic winds, resulting in similar metallicities at the present day \citep[see, e.g., the analytic models by][]{Weinberg2017}.
\citet{Johnson2025} showed that, when ejection is omitted from models of Galactic chemical evolution (GCE), a similar relationship exists between metal yields and the speed of radial gas flows \citep[e.g.,][]{Lacey1985, Spitoni2011, Bilitewski2012}.
This outcome indicates that this degeneracy arises because GCE models require newly produced metals to be mixed into some amount of fresh hydrogen, which is affected by both ejection and radial gas flows \citep{Johnson2025solo}.
\citet{Johnson2025} pointed out that \het/\hef\ and D/H, like metallicity, are indicators of how much of the ISM has undergone nuclear reactions in stellar interiors and could therefore be useful in breaking this degeneracy empirically. \citet{WellerWeinberg2026} have recently highlighted that BBN is the dominant source of \het\ in the Milky Way, and investigated a wide range of parameters to bring models of galactic chemical evolution in line with tight observational measurements. Overall, these studies have highlighted the importance of using \heiso\ and D/H to study the properties of outflows in the Milky Way.

In this paper, we report two new measurements of \heiso\ in the Milky Way and an improved measurement of \heiso\ towards \tet\ \citep{Cooke2022}. In Section~\ref{sec:obs}, we describe the properties of the archival data and the new observations that we analyse herein, and the data reduction procedures adopted. In Section~\ref{sec:analysis}, we describe our analysis procedure and test the variability of \HeIs\ absorption lines. In Section~\ref{sec:results} we use our new \heiso\ measurements, in combination with GCE models, to study the production of the helium isotopes in the Milky Way, and provide a new estimate of the primordial helium isotope ratio, (\heiso)$_{\rm P}$. We also report an estimate of the stellar yields scale. Finally, in Section~\ref{sec:conc}, we summarise our main conclusions and suggest avenues for future research.

\section{Observations}
\label{sec:obs}

\citet{Cooke2022} presented observations of $^{3}$\HeIs\ absorption along the line-of-sight towards \tet\ using the Very Large Telescope (VLT) CRyogenic InfraRed Echelle Spectrograph (CRIRES; \citealt{CRIRES2004}). Motivated by this work, we conducted a dedicated search through public VLT archival\footnote{The ESO archive can be accessed from:\\\url{https://archive.eso.org/cms.html}} high resolution optical spectroscopy covering the \HeIs\,$\lambda3889$ and \HeIs\,$\lambda3188$ transitions. Our search has uncovered many hitherto unrecognised \HeIs\ absorbers in the Milky Way, and highlighted that regions of high gas density and ionization parameter are essential to detect metastable helium absorption lines.

Based on these optical data, we generated synthetic CRIRES observations to test the feasibility of significantly detecting $^{3}$\HeIs\ absorption. We then acquired new observations of our most promising sightlines where \heiso\ could be measured with a modest allocation of telescope time. In this paper, we report our three cleanest detections of $^{3}$\HeIs\ absorption, based on new UVES and CRIRES observations of the stars
\tet,
HD\,319718,\footnote{Note that HD\,319718 is a double star, and one of these stars is in a tight binary.} and
Her 36.

To measure \heiso, we require the combination of: (1) a weak \HefIs\ absorption line, usually the metastable lines at 3188\,\AA\ or 3889\,\AA; and (2) the weak \HetIs\ absorption line at 1.083\,$\mu{\rm m}$. This \HetIs\ absorption line exhibits an isotope shift of $\sim36.6~{\rm km~s}^{-1}$ relative to its companion \HefIs\ line, which is stronger by a factor of $\sim10^{4}$. This measurement also requires a high spectral resolution ($R\gtrsim40,000$) to resolve the line profile, exceptionally high S/N in the near-infrared to detect the usually weak \HetIs\ absorption, and two separate instruments to cover the broad wavelength range. Below, we describe the observations and data reduction procedures that we have adopted for the two instruments that were used in this work. A journal of the observations is provided in Table~\ref{tab:observations}.

\begin{table*}[]
    \centering
    \caption{Journal of observations}
    \begin{tabular}{lcccccccc}
       \hline
       Star name  & RA & Dec & Date  & Instrument$^{\rm a}$ & Slit width & Exposure & S/N$^{\rm b}$ & Programme ID$^{\rm c}$ \\
         & (J2000) & (J2000) &  observed &  & (arcsec) & time (s) & pix$^{-1}$ & \\
      \hline

       \tet        & 05$^{\rm h}$35$^{\rm m}$22.\!\!$^{\rm s}$90 & $-$05$^{\circ}$24$'$57.\!\!$''$83 & 2021 Sep 18 & VLT/CRIRES$^{+}$  & 0.2 & 2040 & 995 & \textbf{107.22U1.001} \\
                   &  &  & 2022 Oct 24 & VLT/CRIRES$^{+}$  & 0.2 & 2090 & 1637 & \textbf{110.23QS.002} \\
                   &  &  & 2014 Sep 24 & VLT/UVES (346) & 0.4 & 1485$^{\rm d}$ & 182 & 194.C-0833(A) \\
                   &  &  & 2022 Oct 3 & VLT/UVES (346) & 0.7 & 39$^{\rm d}$ & 186 & \textbf{110.23QS.001} \\
       HD\,319718  & 17$^{\rm h}$24$^{\rm m}$43.\!\!$^{\rm s}$49 & $-$34$^{\circ}$11$'$57.\!\!$''$03 & 2024 May 13 & VLT/CRIRES$^{+}$ & 0.2 & 4320 & 885 & \textbf{112.25D8.001} \\
                   &  &  & 2009 Jun 11 & VLT/UVES (437) & 0.8 & 1800 & 78 & 081.D-0854(B) \\
                   &  &  & 2024 Mar 20 & VLT/UVES (346) & 1.0 & 2700 & 26 & \textbf{112.25D8.002} \\
       Her 36      & 18$^{\rm h}$03$^{\rm m}$40.\!\!$^{\rm s}$32 & $-$24$^{\circ}$22$'$42.\!\!$''$86 & 2024 Sep 22,30 & VLT/CRIRES$^{+}$ & 0.2 & 5760 & 585 & \textbf{112.25D8.001} \\
                   &  &  & 2013 Jun 10, Sep 4 & VLT/UVES (390) & 0.4 & 10\,435 & 190 & 091.C-0851(A) \\
                   &  &  & 2017 May 3, Aug 20 & VLT/UVES (437) & 0.4 & 3600 & 103 & 194.C-0833(D) \\
                   &  &  & 2024 Mar 20 & VLT/UVES (346) & 1.0 & 2700 & 106 & \textbf{112.25D8.002} \\
      \hline
    \label{tab:observations}

    \end{tabular}
    Notes --- $^{\rm a}$ In parentheses, we list the grating of these observations.\\
    $^{\rm b}$ Signal-to-noise ratio per pixel in the continuum near the \HeIs\ transition wavelength.\\
    $^{\rm c}$ Programme IDs listed in boldface are part of our own dedicated \heiso\ survey.\\
    $^{\rm d}$ Note that the 2014 UVES observations of \tet\ were acquired in $>1.\!\!''8$ seeing with a narrow slit, while our newer 2022 UVES observations were taken in $0.\!\!''6$ seeing with a slit that was well-matched to the seeing. The 2022 observations of 39\,s therefore provide a comparable quality to the 2014 observations.\\

\end{table*}

\subsection{UVES data}
\label{sec:uves}

Although archival optical data exist of these sightlines, \citet{Cooke2022} cautioned that the lifetime of the metastable state of helium is just 130~min, and the available data indicated a possible time variation of the \HefIs\ column density towards \tet. We have therefore acquired new high spectral resolution optical observations of our three target stars to test if the helium metastable state is in ionization equilibrium, or if it slowly varies with time. For this purpose, we observed our three target stars with the Ultraviolet and Visual Echelle Spectrograph (UVES; \citealt{Dekker2000}), which is mounted at the Nasmyth focus on VLT UT2. For the analysis in this paper, we use a combination of both archival and new observations.

The UVES data analysed in this paper were acquired with a range of slit widths from $1.\!\!''0-0.\!\!''4$, resulting in a nominal spectral resolution $R\simeq40\,000-80\,000$ (equivalent to a velocity full-width at half maximum, $v_{\rm FWHM}\simeq3.7-7.2\,{\rm km~s}^{-1}$), respectively. We empirically determine the instrumental line spread function (LSF) from exposures of a ThAr wavelength calibration frame, assuming that each line is unresolved at the UVES spectral resolution, and well-approximated by a single Gaussian profile. In order to cover the requisite 3188\,\AA\ and 3889\,\AA\ absorption lines, we required two configurations. To cover the 3188\,\AA\ line, we used the $346$ grating in combination with the HER\_5 filter, yielding a wavelength coverage $3040-3860$\,\AA. For the 3889\,\AA\ line, we used either the 390 or 437 grating in combination with the HER\_5 filter, yielding a wavelength coverage $3290-4515$\,\AA\ or $3745-4990$\,\AA, respectively. In all cases, we adopted $1\times1$ on-chip binning, and obtained a minimum of three exposures per target to remove cosmic rays. The total exposure times of each target and setup are compiled in Table~\ref{tab:observations}.

The reduce these data, we added UVES data reduction support to the PypeIt spectroscopic data reduction pipeline \citep{PypeIt}.\footnote{PypeIt is available from:\\ \url{https://pypeit.readthedocs.io/en/release/}} The only dataset that are not reduced with PypeIt are the 2022 UVES data of \tet; we instead used \texttt{UVES\_headsort} \citep{Murphy2016} in combination with the standard \texttt{ESOREX} recipes, since these data were acquired with an image slicer. Both pipelines perform a similar set of steps to reduce the data, including an overscan and bias subtraction, followed by orienting and trimming the raw frames. A flat-field calibration was used to account for pixel-to-pixel sensitivity variations, and correct for the spatial illumination profile. A global wavelength calibration solution was constructed with reference to a ThAr lamp, which was also used to trace the spectral tilt of each spectral order. A global two-dimensional (2D) fit was performed to both the wavelength calibration, and the spectral tilt. The spectrum of the science target was optimally extracted, and placed onto a vacuum barycentric\footnote{The barycentric wavelength scale differs by $\lesssim 0.1$\,km~s$^{-1}$ from the more commonly used heliocentric scale.} wavelength scale. All individual exposures were combined using the \textsc{UVES\_popler} software \citep{Murphy2019}, which resampled the data to a constant velocity wavelength grid and allowed us to visually inspect and reject significantly deviant pixels.

\subsection{CRIRES data}
\label{sec:crires}

In 2021, the CRIRES spectrograph was upgraded with three new sensitive Hawaii 2RG detectors and its wavelength coverage was improved by installing a cross-disperser \citep[the instrument is now called CRIRES$^{+}$]{CRIRESp}. CRIRES$^{+}$ is a near-infrared echelle spectrograph, covering the wavelength range from 0.95 to 5.3\,$\mu$m and a spectral resolving power of either $R\simeq40,000$ or $R\simeq80,000$. CRIRES$^{+}$ is installed at the Nasmyth focus of UT3, and can be used in combination with the Multi-Applications Curvature Adaptive Optics (MACAO) system.

\subsubsection{Observations}

As part of the science verification of CRIRES$^{+}$, we acquired observations of \tet\ on 2021 Sep 18, as previously reported by \citet{Cooke2022} (a total integration time of 2040\,s). As described further below, we have improved our data reduction algorithms and re-reduced these data. We supplement these data with new observations acquired on 2022 Oct 24. Our observations were scheduled at specific times of the year so that the heliocentric motion of the Earth introduces a Doppler shift that imprints the weak $^{3}$\HeIs\ absorption line away from telluric lines due to Earth's atmosphere.

We used an identical instrument setup and observing strategy to our 2021 observations. Specifically, we used the MACAO adaptive optics combined with a $0.\!\!''2$ wide slit, producing a nominal instrument resolution of $R\simeq80,000$ ($v_{\rm FWHM}\simeq3.75\,{\rm km~s}^{-1}$). To reduce the impact of pixel-to-pixel sensitivity corrections limiting the final combined S/N, we adopted a non-standard strategy to combine pairs of exposures at different nod positions. For example, an exposure with a $\pm1.\!\!''0$ nod either side of the centre of the slit was acquired adjacent to a nod with $\pm6.\!\!''5$ either side of the centre of the slit. In our data reduction, we then subtracted the A nod from one sequence (e.g. $+1.\!\!''0$) with the B nod from another sequence (e.g. $-6.\!\!''5$) so the A$-$B difference frame contains a positive and a negative trace that are always $\sim7.\!\!''5$ from each other. Other combinations (e.g. $\pm1.\!\!''5$ and $\pm6.\!\!''0$) were also used to maintain the $7.\!\!''5$ difference. This strategy ensures that different pixels on the detector are illuminated with each exposure, and we can confidently subtract any residual background emission, sky emission, or dark current present in our exposures.

For our new \tet\ observations, we acquired twenty exposures with a detector integration time \texttt{DIT}=5\,s and the number of \texttt{DIT}s in one exposure is \texttt{NDIT}=20. We also acquired two exposures with \texttt{DIT}=15\,s and \texttt{NDIT}=3, resulting in a total integration time of 2090\,s.

In addition to the \tet\ observations, we acquired new observations of HD\,319718. The data reported here were acquired on 2024 May 13, and consisted of the same instrument setup described above. We obtained 24 exposures and adopted \texttt{DIT}=180\,s and \texttt{NDIT}=1, resulting in a total integration time of 4320\,s.\footnote{We also obtained observations of this target on 2024 Jul 16 and 2024 Aug 25-27; however, due to a miscalculation of the \HeIs\ heliocentric velocity (based on the original optical data), the $^{3}$\HeIs\ is badly blended with telluric absorption. With a goal to obtain clean and high-quality measures of the \heiso\ ratio, we therefore decided to not include these data in our analysis.} Finally, we also obtained new observations of Her 36, using the same strategy described above. The data were acquired on 2024 Sep 22 and 30, and consisted of 24 total exposures, each with \texttt{DIT}=240\,s and \texttt{NDIT}=1, resulting in a total integration time of 5760\,s.

\subsubsection{Data reduction}
\label{sec:criresdr}

To achieve a high S/N ratio, we wrote a set of custom reduction scripts for each object to reduce the data.\footnote{Although these scripts are not general purpose, we make them available at the following repository (tagged as version v1.0.0-alpha):\\ \url{https://github.com/rcooke-ast/CRIRESredux}} We adopt several standard processing steps, including a dark current subtraction and a flatfield correction to remove the pixel-to-pixel sensitivity variations. Our differencing procedure removed hot pixels, and the CRIRES$^{+}$ detector pattern structure. A Fabry-Perot frame was used to trace the spectral tilt (i.e. pixels of constant wavelength). We used the \texttt{PypeIt} data reduction routines \citep{PypeIt} to trace the object and perform an optimal extraction.

The A$-$B nod strategy that we have adopted is effective at removing the sky emission lines. Therefore, the background regions primarily consist of emission lines from the \HII\ region and any minor residuals from the imperfect sky emission line subtraction. We perform an iterative fit to the background regions and the object profile. The object profile is calculated with a b-spline fit that accounts for the spectral tilt. The optimal extraction is then based on a combination of the object profile and a simultaneous linear fit to the background regions. This is iterated 10 times, after which the background regions are relatively stable and the residuals of the optimal extraction are close to the Poisson limit, showing relatively little structure. As part of this procedure we noticed that the object profile, which is largely based on the continuum regions and accounts for the spectral tilt, is significantly different in regions where the flux has a significant gradient in the spectral direction. We were unable to unambiguously determine the cause of the $10-40$\% changes to the width (broader) and shape (asymmetry) of the object profile in regions of lower flux and higher spectral gradient. Our tests indicate that the spectral tilt inferred from the Fabry-Perot plays a minor role. We also conclude that it is unlikely to be astrophysical in nature, since the same effect is seen with the telluric lines (albeit, less severe). To compensate for this uncertainty, in regions where the spectral gradient is large and we detect significant deviations from the `optimal' profile, we replace the inner $\sim10$ pixels of the optimal profile with the actual detected flux. This implementation introduces some uncertainty to the true object profile, given the noise of the data, and therefore we increase the error of these pixels by a factor of $\sqrt{2}$. We note that this affects the highest flux gradient regions of the \HefIs\ absorption line, and not the regions where we detect the $^{3}$\HeIs\ absorption. We therefore do not include the affected pixels in our line profile fitting.

After the data were optimally extracted, we obtained a first guess of the wavelength solution by simultaneously fitting the \HeIs\,1.083\,$\mu{\rm m}$ triplet absorption of all individually extracted CRIRES$^{+}$ spectra at the same time as fitting the optical data (see Section~\ref{sec:uves}), which are registered onto a vacuum barycentric wavelength scale. This global fit also determines a continuum level of all the exposures, consisting of a Legendre polynomial that is the same for all exposures, and a continuum scale and tilt for each individual exposure to account for small differences in wavelength dependent sensitivity between each exposure. The order of the global Legendre polynomial was increased until the residuals were not significantly correlated on large scales. Given that the optical data provide a ground truth reference to the vacuum barycentric wavelength scale, we include two free parameters per exposure to allow for a linear transformation from pixels to wavelength over the $\sim10$\,\AA\ wavelength interval around the absorption line. For further details about the fitting procedure, see Section~\ref{sec:analysis}. The relative scale and tilt of the continuum, combined with the linear transformation of the wavelength scale, are used as a reference to combine all individual exposures into a single combined spectrum with a pixel sampling of $1.5\,{\rm km~s}^{-1}$ (i.e. sampling the FWHM of the LSF with about 2.5 pixels). Based on the fluctuations in the continuum regions free of absorption in the blue wing of the profiles, we estimate a final combined signal-to-noise ratio (S/N) per $1.5\,{\rm km~s}^{-1}$ pixel of
S/N=995 and 1637  for \tet\ (2021 and 2022 data, respectively),
S/N=885 for HD\,319718, and
S/N=585 for Her 36. Note that the S/N of our data are not limited by the flatfield frames. We obtained $25\times7~{\rm sec}$ flatfield frames, each with a S/N per detector pixel S/N=305. Our observation strategy projects the target light nearly uniformly onto $\sim120$ detector pixels. Therefore, the effective S/N of the flatfield is $\sim3300$, which is much higher than the S/N of our target spectra.

\section{Analysis}\label{sec:analysis}

We use the Absorption LIne Software (\textsc{alis}) package\footnote{\textsc{alis} can be downloaded from: \url{https://github.com/rcooke-ast/ALIS}. The v1.0.0-alpha tagged version of the code was used in this paper.} to model the stellar continuum as well as the absorption line profiles from the intervening gas. \textsc{alis} employs a Levenberg-Marquardt algorithm to find the optimal model parameters that minimize the $\chi^{2}$ statistic. For our analysis, we perform a simultaneous fit to the stellar continuum, the \HeIs\ absorption, the wavelength solution, and the telluric absorption. By modelling all of these parameters simultaneously, we account for the dominant systematic uncertainties that might affect the determination of \heiso, and these uncertainties will be incorporated into the determination of the model parameters.

\subsection{Variability of metastable absorption}
\label{sec:timevary}

For the first step of our analysis, we use two time-separated measurements of the \HefIs\ absorption along the line-of-sight to \tet\ to measure the time-evolution of the strength of the absorption line profile. There are three reasons why this test is important: (1) the metastable helium state has a lifetime of just 130~min; if the gas cloud departs from radiative equilibrium, significant changes to the metastable absorption might be expected on the timescales of minutes to hours; (2) the absorbing medium may exhibit a transverse motion and either structure or gradients perpendicular to the line-of-sight. This may cause the column density of the absorption to be different during two time-separated epochs \citep[e.g.][]{Boisse2015}; and (3) it is important to know if simultaneous observations are required to accurately measure the helium isotope ratio.

Until now, the only assessment of the time variability of metastable helium absorption was reported by \citet{Cooke2022}, who found a significant systematic difference between the column density of \HefIs\ derived from weak optically thin absorption lines in the optical wavelength range compared to optically thick absorption measured at near-infrared wavelengths. Other than intrinsic variability of the absorption line column density, other explanations for this disagreement include the wavelength calibration or radiative transfer effects that could alter the shape of the absorption line profile relative to the weaker absorption lines. We consider these alternative hypotheses to be more likely, given that the strength of the \HeIs\ absorption has remained remarkably constant for almost 100 years \citep{Wilson37}.

The optimum way to measure the time-variability is to use the same optically thin line from well-separated epochs, ideally with the same instrument and setup. We therefore simultaneously analyse archival and new observations of the \HeIs\,$\lambda3188$\,\AA\ absorption line, separated by just over eight years ($\Delta t=8.02486~{\rm yrs}$). We model this absorption with a single Voigt profile, consisting of a column density, a relative velocity offset, and a Doppler parameter. We allow for a velocity offset between epochs to account for uncertainties in the wavelength calibration. The absorption profile observed at each epoch has the flexibility to have a different column density and Doppler parameter. The stellar continuum is well-modelled with a seventh order Legendre polynomial. We fit directly to the rate of change of the logarithmic column density, ${\rm d}\log_{10}\,[N({\rm He\,\textsc{i}}^{*})/{\rm cm}^{-2}]/{\rm d}t$, and rate of change of the Doppler parameter, ${\rm d}b/{\rm d}t$. We obtain an excellent fit to the data (with a reduced chi-squared $\chi^{2}/{\rm dof}$=0.831) with a best-fitting value ${\rm d}\log_{10}\,[N({\rm He\,\textsc{i}}^{*})/{\rm cm}^{-2}]/{\rm d}t = (2.4 \pm 2.4)\times10^{-4}~{\rm dex~yr}^{-1}$, and ${\rm d}b/{\rm d}t = (3.5 \pm 3.8)\times10^{-3}~{\rm km~s^{-1}~yr^{-1}}$. Given that these values are consistent with no variation, we therefore report an upper limit on the time-variability of metastable helium absorption in the Orion nebula of ${\rm d}\log_{10}\,[N({\rm He\,\textsc{i}}^{*})/{\rm cm}^{-2}]/{\rm d}t\leq7.2\times10^{-4}~{\rm dex~yr}^{-1}$ ($2\sigma$ confidence), which is equivalent to $\leq0.17\,\%~{\rm yr}^{-1}$ when expressed as a linear quantity. Similarly, we report an upper limit on the possible variation of the Doppler parameter, ${\rm d}b/{\rm d}t \leq 11~{\rm m~s^{-1}~yr^{-1}}$ ($2\sigma$ confidence). For a pure thermally broadened model, this upper limit on the broadening corresponds to an upper limit on the temperature variation ${\rm d}T/{\rm d}t \leq 36~{\rm K~yr^{-1}}$ ($2\sigma$ confidence). In the subsequent analysis, we assume that the column density of metastable helium absorption does not change substantially over the timescales of our observations.\footnote{It is possible that our observations were acquired when the \HeIs\ column densities were \emph{measured} to be nearly identical, even though there is some underlying time variability. We appeal to Occam's Razor to rule out this possibility, and we note that simultaneous observations are ideally suited to ensure the optical and near-infrared absorption lines arise from the same sightline.} Since our most recent UVES and CRIRES$^{+}$ observations were acquired within a few weeks to months of each other (see Table~\ref{tab:observations}), we conclude that it is unlikely the inferred \heiso\ ratios reported here are affected by time variability of \HeIs. In the future, simultaneous observations with a single instrument covering a wavelength range of 3180\,\AA--10900\,\AA, or two separate instruments covering the near-ultraviolet and the near-infrared operated simultaneously, would be preferred to avoid this assumption.

\subsection{Profile fitting}
\label{sec:fitting}

As noted by \citet{Cooke2022}, the optically thick \HeIs\ absorption lines appear asymmetric and not well-modelled by a single (symmetric) Voigt profile at the very high S/N of our data. With the new data reported here, we have also noted that the optically thick \HeIs\ absorption lines are blue-shifted relative to the optically thin transitions; for \tet, we detect a significant blueshift of $1.82\pm0.23~{\rm km~s}^{-1}$. These indications suggest that the optically thick gas responsible for \HeIs\ absorption is expanding. To appropriately model the line profiles of the \emph{optically thick} lines, we should perform radiative transfer calculations of a thin slab of expanding dense gas exposed to an incident radiation field with a high ionization parameter.

Fortunately, the primary goal of our observations is to measure the \het/\hef\ ratio. Given the time-stability of weak optically thin metastable helium absorption (see Section~\ref{sec:timevary}), we only need to model the optically thin lines, and we have found these to be well-modelled with Voigt profiles. The gas is therefore well-approximated by a Maxwellian distribution of velocities. A single Voigt profile consists of three free parameters, including a column density that governs the overall strength of the absorption line, a velocity offset relative to the Solar System barycentre, and a Doppler parameter that governs the width of the absorption.

The Doppler parameter of an ideal absorber consists of two comopnents: (i) a turbulent component ($b_{\rm turb}$) that describes the macroscopic motions of the gas, and is the same for all ions; and (ii) a thermal component that accounts for the microscopic motions of the gas. The total broadening is given by:
\begin{equation}
b_{\rm total}^{2} = b_{\rm turb}^{2} + \frac{2\,k_{\rm B}\,T}{m_{\rm ion}}
\end{equation}
where $k_{\rm B}$ is the Boltzmann constant, $T$ is the temperature of the absorbing gas, and $m_{\rm ion}$ is the mass of the absorbing ion. Given that \het\ and \hef\ have a different atomic mass, we can in principle decouple the contributions of turbulent and thermal broadening by comparing the relative widths of the \het\ and \hef\ line profiles. As discussed further in Section~\ref{sec:tet}, we find that our simplest absorber (\tet) prefers a model where $b_{\rm turb}$ is consistent with zero, and the line profile is dominated by thermal broadening. We also find that pure thermal broadening fits to each absorber yield a lower reduced $\chi^{2}$-statistic than a pure turbulent model. We therefore assume that the line profiles are purely broadened by the thermal motions of the absorbing gas, and set the turbulence to zero (i.e. $b_{\rm turb}=0~{\rm km~s}^{-1}$).

Some of the sightlines presented here intersect multiple absorbing `clouds' along the line-of-sight, and each cloud is modelled as a Voigt profile. We assume that each absorbing cloud has the same intrinsic \het/\hef\ ratio; we therefore fit directly to the linear column density ratio $N(^{3}$\HeIs)/$N(^{4}$\HeIs). Given the nearly identical ionization energies of \het\ and \hef, we also assume that charge transfer reactions ensure that $N(^{3}$\HeIs)/$N(^{4}$\HeIs) closely follows the intrinsic helium isotope ratio, \heiso~=~$N(^{3}$\HeIs)/$N(^{4}$\HeIs).

The \HeIs\,$\lambda1.0833\,\mu{\rm m}$ transition is proximate in wavelength to several unrelated telluric absorption lines that are formed in Earth's atmosphere. We model these telluric absorption features with Voigt profiles, but include an additional damping term (the same value for all telluric lines) to model the collisional broadening of the line profile. Finally, we note that the \het\ absorption lines exhibit hyperfine structure. As justified in \citet{Cooke2022}, we assume the relative level population of the $F=0.5$ and $F=1.5$ hyperfine levels is given by the ratio of the level degeneracies (i.e. $n_{F=0.5}/n_{F=1.5}=0.5$). We adopt the atomic data collated in Table~1 of \citet{Cooke2022}.

In all cases, we model the stellar continuum with low order Legendre polynomials (typically of order 5, where the polynomial order is chosen to minimize correlated residuals), where the coefficients of these polynomials are fit simultaneously with the \HeIs\ absorption lines so that the continuum uncertainty is included in our reported uncertainties on all model parameters. One \HeIs\,$\lambda3188$ absorption line is used as a reference for the systemic velocity of the system. For all other optical absorption lines, we include a constant velocity shift as a model parameter to account for uncertainties in the wavelength calibration that might be present when  ${\rm S/N}\gtrsim100$. For the near-infrared CRIRES$^{+}$ data covering the \HeIs\,$\lambda1.0833\,\mu{\rm m}$ transition, we instead include a shift and stretch correction to the wavelength calibration during the fitting procedure. The shift and stretch correction accounts for linear deviations from our initial linear transformation from pixels to wavelength (see Section~\ref{sec:criresdr}).

\subsection{Individual Systems}
\label{sec:systems}

All systems were consistently analysed using the approach described in Section~\ref{sec:fitting}, however, different systems contain a different number of absorption components. Each system was independently modelled, and in this section we describe the individual properties of the absorption line systems presented in this work.

\subsubsection{\tet}
\label{sec:tet}

\begin{figure*}
\includegraphics[width=\textwidth]{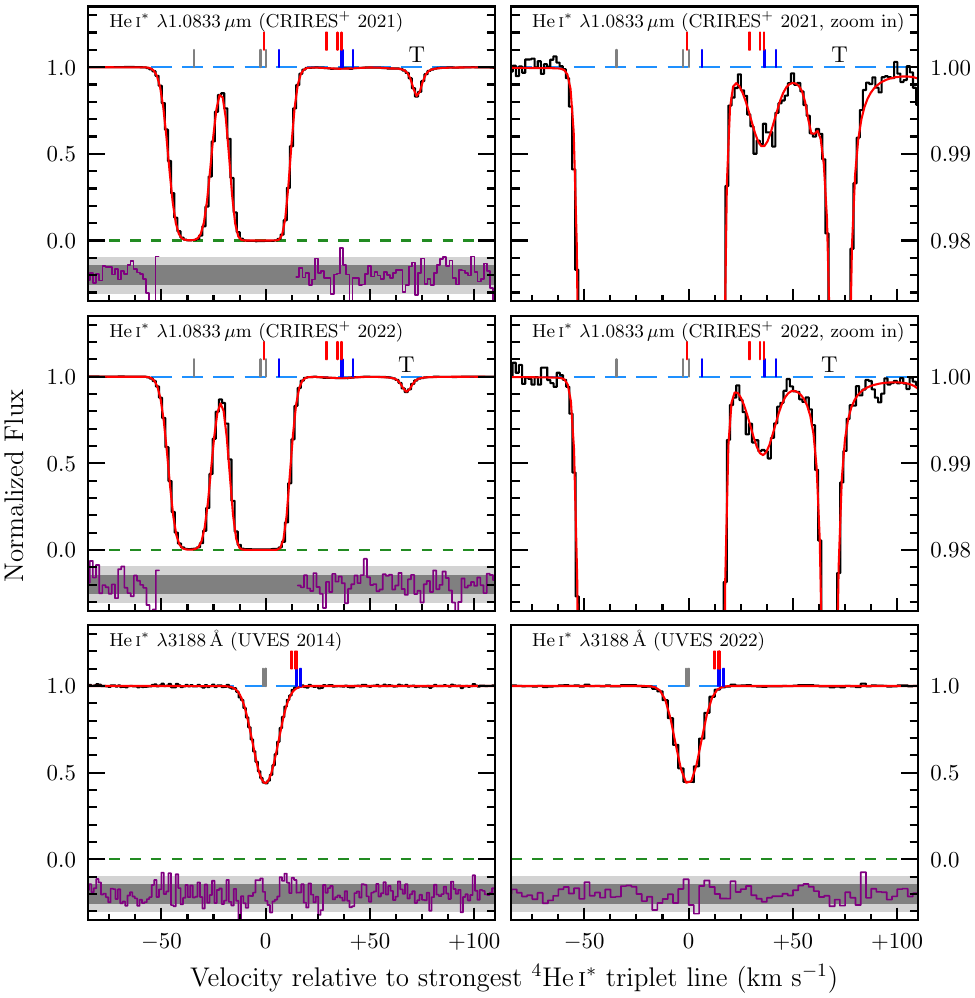}
\caption{\HeIs\ absorption lines along the sightline to \tet\ (black histograms) are overlaid with a best-fitting model (red curves). The (data$-$model)/error residuals are shown at the bottom of each panel (purple histograms), where the 68 and 95 percent confidence intervals are shown by the dark and light shaded bands, respectively. The top and middle panels show our CRIRES$^{+}$ data, while the bottom panels show the UVES data. The top right and middle right panels show a zoom-in of the top left and middle left panels, respectively. The tick marks above the spectra show the locations of the \HefIs\ absorption components (grey ticks) and \HetIs\ absorption  components (blue and red ticks indicate the $F=0.5$ and $F=1.5$ levels, respectively). Note that we only display the tick marks for the primary component responsible for the observed \HetIs\ absorption. Several telluric lines are present in the CRIRES$^{+}$ spectra, and are marked with a `T'. We have not included the optically thick \HefIs\,$\lambda1.0833~\mu{\rm m}$ absorption in the final fits (see Section~\ref{sec:fitting} for further details), but the red model is still displayed for comparison. The reduced chi-squared of this fit is $\chi^{2}/{\rm dof}=1.246$.
\label{fig:modelfits_tet}}
\end{figure*}

The absorption cloud along the line-of-sight to \tet\ is both the simplest and the strongest \HeIs\ absorption line system that we have analysed. We find that a single Voigt profile is sufficient to model the absorption line system. We have not included the \HeIs\,$\lambda3889$\,\AA\ absorption line in the model fits, since this line is close to saturation, and is possibly affected by line profile asymmetries due to radiative transfer effects that are not accounted for. Given the \HeIs\,$\lambda3188$\,\AA\ line is symmetric, optically thin, invariant over the $\sim8$ year time interval of our data, and consistent with a single Voigt profile, we have therefore used this absorption line (from both epochs) as the primary measure of the \HefIs\ column density. \HetIs\ absorption is also imprinted on the \HeIs\,$\lambda3188$\,\AA\ profile, but it is extremely weak and only exhibits an isotope shift of $\sim 15.5~{\rm km~s}^{-1}$. The dominant line that determines the total column density of \HetIs\ is the \HeIs\,$\lambda1.0833\,\mu{\rm m}$ absorption line. Nevertheless, we simultaneously include both \het\ and \hef\ absorption lines in all transitions of the available data.

Since this absorption line system exhibits a profile that is consistent with a single absorption component, we used the data of this absorber to separate the contributions of thermal and turbulent broadening, based on the relative line widths of \het\ and \hef. The $\chi^{2}$ minimization procedure favoured a solution with zero turbulence, implying that the thermal motion of the absorbing ions is sufficient to reproduce the data. In what follows, we have therefore set the turbulent broadening to zero.

Figure~\ref{fig:modelfits_tet} shows the data and best-fitting model, corresponding to a reduced chi-squared $\chi^{2}/{\rm dof}=1.246$. Overall, the model is an acceptable fit to the data, although we do note that the residuals are correlated for the \HeIs\,$\lambda3188$\AA\ absorption. This could be explained if the LSF of UVES at this wavelength deviates from a Gaussian, or if the intrinsic line profile departs from the assumed Voigt profile. The total \HefIs\ column density of the absorption line is
$\log_{10}(N(^{4}{\rm He}\,$\textsc{i}$^{*})/{\rm cm}^{-2})=13.6943\pm0.0014$, while the best-fitting kinetic temperature of the gas is
$T=19820\pm150~{\rm K}$. The best-fitting value of the helium isotope ratio of this sightline is:

\begin{equation}
    \Theta^{2}{\rm A~Ori}:\quad
    \frac{N(^{3}\mathrm{He\,\mathsc{i}}^{*})}{N(^{4}\mathrm{He\,\mathsc{i}}^{*})} = (1.855\pm0.067)\times10^{-4}
\end{equation}

This corresponds to a 3.6 percent determination of the helium isotope ratio of the Orion Nebula. The central value agrees with that reported by \citet[][$^{3}{\rm He}/^{4}{\rm He}=(1.77\pm0.13)\times10^{-4}$]{Cooke2022}, but with a factor of $\sim 2$ improved precision mainly thanks to the improved S/N of the new CRIRES$^{+}$ data. The slightly higher value is due to our decision to fit a pure thermally broadened Voigt line profile in this work instead of a pure turbulent spline broadening profile, as used by \citet{Cooke2022}. In our previous work, we only modelled the optically thick \HefIs\,$\lambda1.0833\,\mu{\rm m}$ absorption line, and this required us to adopt a spline optical depth profile that was not designed to model the effects of thermal broadening. We have now found that the \HeIs\ column density does not change significantly with time (Section~\ref{sec:timevary}), so we can therefore adopt a Voigt profile model, including the effects of thermal broadening, and fit only the weak optically thin lines. We note that the column density of \HefIs\ that we report here is consistent with that derived from the optical lines reported by \citet{Cooke2022}. From the \emph{Gaia} coordinates and parallax measurement of \tet, we infer that the \HeIs\ absorber along this sightline is located at a galactocentric radius $R_{\rm gal}=8.4~{\rm kpc}$.

\subsubsection{HD\,319718}

\begin{figure*}
\includegraphics[width=\textwidth]{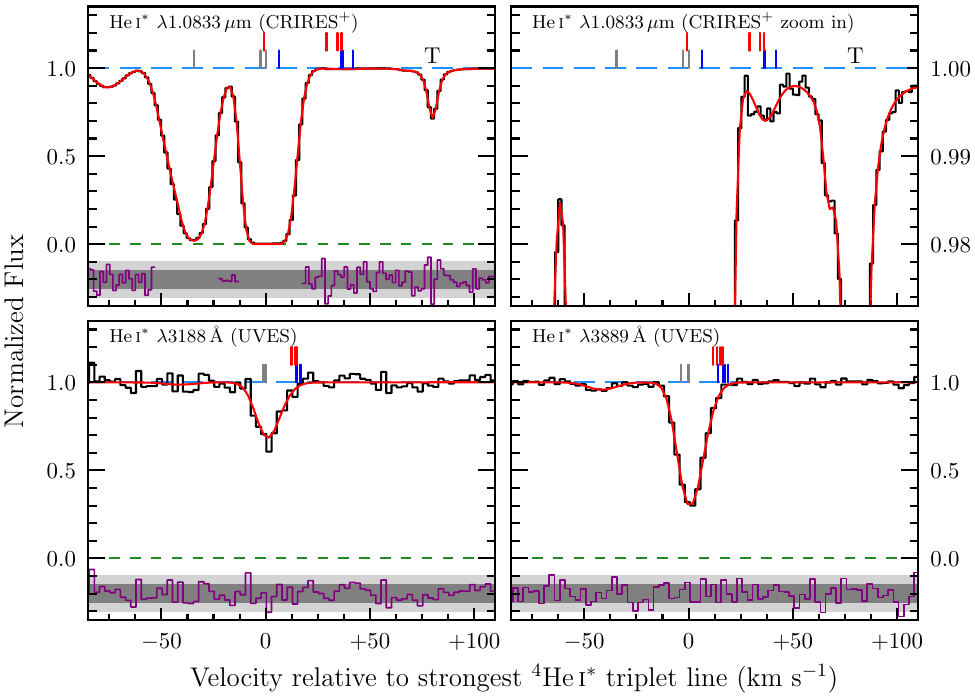}
\caption{Same as Figure~\ref{fig:modelfits_tet}, but for the sightline towards HD\,319718. Note that we only display the tick marks for the primary component responsible for the observed \HetIs\ absorption. The reduced chi-squared of this fit is $\chi^{2}/{\rm dof}=1.520$.
\label{fig:modelfits_hd}}
\end{figure*}

The designation HD\,319718 (Pismis 24-1) comprises at least three stars, including two physically unrelated O stars of comparable magnitude separated by $\sim0.\!\!''36$ in projection on the sky based on \emph{Hubble Space Telescope} imaging observations \citep{Maiz2007}. Radial velocity variations of the stellar absorption lines indicate that one of these stars is an unresolved spectroscopic binary with a period of $\sim2.4$ days \citep{Maiz2007,Barr2013}. Since our optical data unavoidably include the light from all stars belonging to HD\,319718, our CRIRES$^{+}$ spectra are also designed to observe the two brightest stars associated with HD\,319718, but with the advantage of adaptive optics. From our CRIRES$^{+}$ data, we measure a separation of $\sim0.\!\!''36$, in good agreement with the \emph{HST} data, and find that the \HeIs\ absorption is the same along both sightlines. This indicates that the \HeIs\ absorbing medium is in the foreground of both stars.

To reduce our CRIRES$^{+}$ data, we assume that both stars probe a single sightline through the \HeIs\ absorption, and optimally extract the blended profile assuming it is just one background light source. This ensures that our reduction is consistent with the spatially unresolved stars acquired with our optical observations. We find that the \HeIs\ absorption consists of a main component and a weak satellite absorption component shifted by $\Delta v\simeq-40\,{\rm km~s}^{-1}$ relative to the main component. The main component is well-modelled by two closely separated Voigt profiles ($\Delta v\simeq3\,{\rm km~s}^{-1}$) with column densities
$\log_{10}(N(^{4}{\rm He}\,$\textsc{i}$^{*})/{\rm cm}^{-2})=13.166\pm0.020$
and
$\log_{10}(N(^{4}{\rm He}\,$\textsc{i}$^{*})/{\rm cm}^{-2})=13.011\pm0.027$, while the best-fitting kinetic temperatures are
$T=12550\pm630~{\rm K}$ and
$T=25820\pm340~{\rm K}$, respectively. The weak satellite absorption feature also requires two closely separated Voigt profiles, with a total column density $\log_{10}(N(^{4}{\rm He}\,$\textsc{i}$^{*})/{\rm cm}^{-2})=12.04$. We note that each \HefIs\ component has a corresponding \HetIs\ component, and all components share the same helium isotope ratio. The reduced chi-squared of our best-fit model is $\chi^{2}/{\rm dof}=1.520$, and we determine the helium isotope ratio along this sightline to be:

\begin{equation}
    {\rm HD\,319718}:\quad
    \frac{N(^{3}\mathrm{He\,\mathsc{i}}^{*})}{N(^{4}\mathrm{He\,\mathsc{i}}^{*})} = (2.13\pm0.22)\times10^{-4}
\end{equation}
corresponding to a precision of $\sim11\%$. The best-fit model is shown in Figure~\ref{fig:modelfits_hd}. The \emph{Gaia} coordinates and parallax measurement of HD\,319718 imply that the galactocentric radius of the \HeIs\ absorber is $R_{\rm gal}=6.5~{\rm kpc}$. This absorber therefore offers an important lever arm to infer the scale of the stellar yields, since the regions closer to the centre of the Milky Way are more enriched with the post-BBN yields of \het\ and \hef.

\subsubsection{Her\,36}

\begin{figure*}
\includegraphics[width=\textwidth]{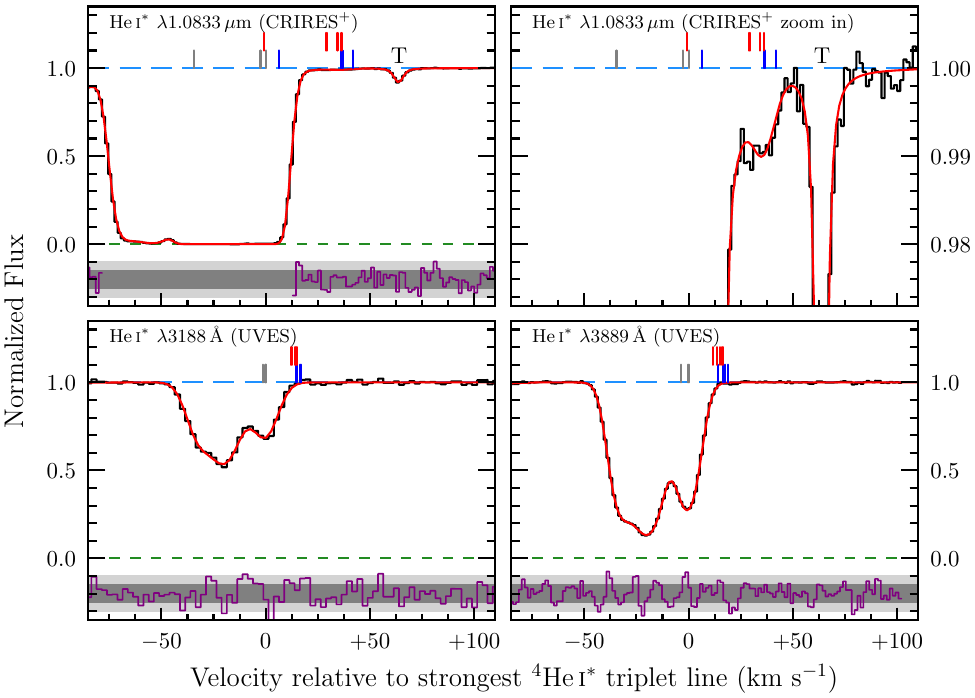}
\caption{Same as Figure~\ref{fig:modelfits_tet}, but for the sightline towards Her 36. Note that we only display the tick marks for the primary component responsible for the observed \HetIs\ absorption. The reduced chi-squared of this fit is $\chi^{2}/{\rm dof}=1.032$.
\label{fig:modelfits_her}}
\end{figure*}

The \HeIs\ line profile imprinted on the spectrum of Her\,36 contains multiple absorption components. We find that the absorption profile is well-modelled ($\chi^{2}/{\rm dof}=1.032$) by three strong components, and two weak satellite components that are blue-shifted by $-86\,{\rm km~s}^{-1}$ and $-100\,{\rm km~s}^{-1}$ relative to the reddest component. For the purposes of this work, we are primarily interested in measuring the helium isotope ratio; we therefore focus this summary on the \emph{reddest} absorption component, where we have detected corresponding absorption due to \HetIs. The \HefIs\ column density of this component is
$\log_{10}(N(^{4}{\rm He}\,$\textsc{i}$^{*})/{\rm cm}^{-2})=13.3902\pm0.0028$, while the best-fitting kinetic temperature is
$T=19780\pm240~{\rm K}$. The best-fitting value of the helium isotope ratio of this sightline is:
\begin{equation}
    {\rm Her\,36}:\quad
    \frac{N(^{3}\mathrm{He\,\mathsc{i}}^{*})}{N(^{4}\mathrm{He\,\mathsc{i}}^{*})} = (3.98\pm0.84)\times10^{-4}
\end{equation}
Despite being of lower precision ($\sim21\%$) and displaying a more complex absorption profile, we detect companion \HetIs\ at the correct wavelength and with an absorption strength comparable to the other two detections reported here. We note that our optical data are not quite sensitive enough to rule out the possibility that this \HetIs\ absorption is blended with a weak satellite \HefIs\ absorption. Nevertheless, given that optical and near-infrared \HeIs\ absorption lines have a different isotope shift \citep{Cooke2022}, future optical data of higher S/N ratio will allow us to identify or rule out the existence of satellite absorption features. The best-fit model is shown in Figure~\ref{fig:modelfits_her}. Based on the \emph{Gaia} coordinates and parallax measurement to Her\,36, we determine that the \HeIs\ absorber is located at a galactocentric radius $R_{\rm gal}=6.9~{\rm kpc}$.

\subsection{Properties of metastable helium absorption}

The three sightlines reported here exhibit some common features. In all cases, despite the high S/N of the spectra, the optically thin absorption lines are consistent with a Voigt profile dominated by thermal broadening. The inferred kinetic temperature of the \HetIs-bearing gas is typically $\gtrsim20,000~{\rm K}$. This supports the idea that the gas is proximate to the \HII\ region, in a thin dense shell with a high ionization parameter. Given that the strengths and widths of the absorption lines do not exhibit time-variability, the gas must be close to radiative equilibrium. We also identified slightly asymmetric profiles for the optically thick lines, that are also slightly blue-shifted relative to the optically thin lines. This indicates that the \HeIs\ absorbing material is expanding with time. We leave a more detailed investigation into the nature of \HeIs\ absorbers to future work.

\section{Results}\label{sec:results}

\het\ has never been detected in an extragalactic environment. Indeed, finding suitable environments in the Milky Way where its abundance can be robustly measured has proven to be challenging. The three measurements that we report here therefore offer a unique opportunity to learn about the primordial and post-BBN production of \het.

Comparing the helium isotope ratio to the metallicity of the absorbing gas allows us to empirically assess its chemical evolution. Unfortunately, we currently do not have the means to measure the metallicity of the absorbing gas cloud, but this may one day become possible with far ultraviolet observations that cover the hydrogen Lyman series and metal absorption lines associated with the absorbing gas (at rest wavelengths between 1100$-$1500\,\AA). For now, we must therefore appeal to alternative approaches to estimate the metallicity of the \het\ absorbers.

In this work, we have assumed that the \HeIs\ absorbing gas is physically close to the background \HII\ region, since it is only these environments that provide the necessary physical conditions (i.e. high gas density and ionization parameter) for helium atoms to reside in the metastable state. The background \HII\ regions provide many diagnostics of the metal abundance, including a direct measure of the O/H abundance from the strong nebular and auroral lines of [O\,\textsc{iii}]. The abundances of the \HII\ regions associated with these stars have previously been reported in the literature, including
\tet: $12+\log_{10}({\rm O/H})=8.51\pm0.03$ (M42; \citealt{ArellanoCordova2020})
HD\,319718: $12+\log_{10}({\rm O/H})=8.39^{+0.15}_{-0.17}$ (NGC 6357; \citealt{Bohigas2004})
Her\,36: $12+\log_{10}({\rm O/H})=8.48\pm0.04$ (M8; \citealt{ArellanoCordova2020}). Thus, within the current errors, the metallicities of these \HII\ regions are consistent with being drawn from a constant value, and compatible with the typically measured slope of the Milky Way radial abundance gradient, $\nabla[{\rm O/H}]\simeq-0.05~{\rm dex~kpc}^{-1}$ \citep{Balser2011,Wenger2019,ArellanoCordova2020,Willett2023,Johnson2025RMG}.

On a similar note, we point out that the measured values of \heiso\ reported here are all comparable, and statistically consistent with being drawn from a constant value. This indicates that the O/H and \heiso\ measurement precisions of our sightlines are not currently sufficient to empirically determine the chemical evolution of \heiso. We have therefore opted to use the galactocentric radius determinations, combined with a model of galactic chemical evolution, to deduce the most likely value of the primordial helium isotope ratio based on our measurements.

\subsection{Galactic Chemical Evolution of \emph{$^{3}$He/$^{4}$He}}
\label{sec:gce}

In this section, we compare our helium isotope ratio measurements to the GCE models by \citet{Johnson2025RMG}, which we integrate using the publicly available \texttt{Versatile Integrator for Chemical Evolution} (\texttt{VICE}; \citealt{Johnson2020}).
In \citet{Cooke2022}, we compared our measurement of \heiso\ in the Orion Nebula to the original forms of these models, presented by \citet{Johnson2021}.
These one-dimensional GCE models discretize the Milky Way disk into 200 concentric rings of width $\delta R = 100$ pc, ranging from $R = 0$ to $20$ kpc, and covering $13.2$ Gyr of star formation.
This ``multi-zone'' framework allows our models to capture the evolution of \heiso\ across the full range of Galactocentric radii.
Each ring has its own star formation rate (SFR) and gas reservoir, which is assumed to be chemically homogeneous.
Neighbouring rings are coupled to one another by the exchange of stellar populations through radial migration (see discussion below), but enrichment otherwise proceeds according to a conventional one-zone GCE model (see, e.g., the reviews by \citealt{Tinsley1980} and \citealt{Matteucci2021}).
\par
These GCE models have previously been used to study abundance patterns of various elements (e.g., nitrogen, \citealt{Johnson2023nitrogen}; carbon, \citealt{Boyea2025}), but we focus on O, \het, and \hef\ in this paper.
Massive stars and their supernovae produce all three, but asymptotic giant branch (AGB) stars also produce \het\ and \hef.
Our yield prescription is taken directly from our recent exploration of these models in \citet{Johnson2025}.
Due to their short lifetimes, \texttt{VICE} injects massive star yields instantaneously after the formation of a stellar population.
The rate of massive star (MS) enrichment of some isotope $x$ can then be expressed as
\begin{equation}
\dot\Sigma_{x}^{\text{MS}} = y_{x}^{\text{MS}} \dot\Sigma_\star,
\end{equation}
where $\dot\Sigma_\star$ is the surface density of star formation, and $y_{x}^{\text{MS}}$ is the population-averaged net yield of $x$ from massive stars, describing the mass of newly produced material per mass of star formation.
\par
We vary the overall normalization of metal yields as a free parameter in combination with (\het/\hef)$_p$.
We refer to the overall scale as $y / Z_\odot$, or the yield relative to solar metallicity.
\citet{Weinberg2024} advocate for the $y / Z_\odot = 1$ scale based on the mean Fe yield of core collapse supernovae measured by \citet{Rodriguez2023}.
At this scale, $y_\text{O,MS} = Z_{\text{O},\odot} = 0.00572$ based on the \citet{Asplund2009} measurements of the composition of the solar photosphere.\footnote{This is consistent with more recent measurements of the solar oxygen abundance \citep{Lodders2021,Pietrow2023}.}
We determine the massive star yield of \hef\ by computing \citeauthor{Weller2025}'s \citeyearpar{Weller2025} recommended total \hef\ yield from their Equation 12, which ensures that the model reproduces the observed relationship between elemental helium and oxygen abundances.
The massive star yield of \het\ is subdominant compared to AGB stars, so we simply retain the \citet{Limongi2018} yield from non-rotating massive stars from our previous work in \citet{Cooke2022}.
\par
We use the \citet{Lagarde2011, Lagarde2012} tables of \het\ and \hef\ yields from AGB stars, which shed their envelopes at a rate that depends on the stellar mass-lifetime relation; we adopt the \citet{Larson1974} mass-lifetime relation.
Our results are not significantly affected by switching to a more recent form \citep[e.g.,][]{Padovani1993, Kodama1997, Hurley2000, Vincenzo2016} because the AGB star enrichment rate is determined by the IMF in the $M \approx 1 - 8~M_\odot$ mass range, where these prescriptions are broadly consistent with one another.
The total AGB star enrichment rate also depends on the initial mass function, for which we adopt the \citet{Kroupa2001} form.
\par
We retain the ``rise-fall'' star formation history (SFH) from previous versions of these models:
\begin{equation}
\label{eqn:taurisefall}
\dot \Sigma_\star \propto \left(1 - e^{-t / \tau_\text{rise}}
\right)
e^{-t / \tau_\text{fall}}.
\end{equation}
We set the normalization of the SFH at each radius such that the model accurately reproduces the stellar mass ($\sim$$5 \times 10^{10}$ M$_\odot$; \citealt{Licquia2015}) and radial density profile of the Milky Way \citep{Bland-Hawthorn2016}.
The values of $\tau_\text{rise}$ and $\tau_\text{fall}$ increase with Galactocentric radius, reflecting the ``inside-out'' growth of the disk \citep[e.g.,][]{Bird2013}.
Their values are chosen such that the model reproduces the observed median stellar age at each radius (see parameter description in \citealt{Johnson2025RMG}).
The SFR and the gas supply are related by a spatially-resolved implementation of the Kennicutt-Schmidt relation, $\dot \Sigma_\star \propto \Sigma_g^N$ where $N = 1.5$ \citep[e.g.,][]{Schmidt1959, Schmidt1963, Kennicutt1998}.
With the prescription for outflows described below, \texttt{VICE} is able to solve the mass continuity equation.
\par
Radial migration \citep[e.g.,][]{Sellwood2002, Loebman2011, Minchev2011} follows the simple, Gaussian random sampling algorithm described in \citet{Dubay2024}, wherein most stars migrate $\sim$$2.7$ kpc by the time they reach an age of $8$ Gyr.
Most importantly, \texttt{VICE} deposits all nucleosynthetic products at the present location of each stellar population, allowing them to enrich distributions of Galactocentric radius as they migrate.
This prescription is particularly important for \het, whose production is dominated by lower mass stars with long lifetimes that may migrate significant distances before shedding their envelopes (see discussion in \citealt{Cooke2022}).
We refer to \citet{Johnson2021} for further discussion of the influence of radial migration on GCE models.
\par
Our models include ejection of ISM gas to the circumgalactic medium (CGM; see, e.g., the reviews by \citealt{Tumlinson2017} and \citealt{Thompson2024}).
The rate of ejection is specified by the mass loading factor, which quantifies the mass loss rate relative to star formation: $\eta_\text{wind} \equiv \dot\Sigma_\text{ej} / \dot\Sigma_\star$.
Following \citet{Johnson2025RMG}, we use an exponential scaling of $\eta_\text{wind}$ with Galactocentric radius:
\begin{equation}
\eta_\text{wind} = \eta_{\text{wind},\odot} e^{(R - R_\odot) / R_\eta}.
\end{equation}
where $R_\eta$ is the scale radius, and the normalizing factor $\eta_{\text{wind},\odot}$ refers to the value of $\eta_\text{wind}$ near the Sun, at $R_\odot = 8$ kpc.
We determine this factor from the scale of stellar yields, according to $\eta_{\text{wind},\odot} = y/Z_\odot - 0.6$, in order to ensure that the predicted abundances remain empirically plausible at all choices of $y / Z_\odot$.
\par
The increase in $\eta_\text{wind}$ establishes the radial metallicity gradient \citep[e.g.,][]{Maiolino2019, Sanchez2020} in the Galactic disk at low redshift in these models.
We use a scale radius of $R_\eta = 7$ kpc throughout, which corresponds to a metallicity gradient slope near $\nabla$[O/H] $= -0.06$ kpc$^{-1}$ (see discussion in \citealt{Johnson2025RMG}).
Some GCE models \citep[e.g.,][]{Grisoni2018, Palla2020} use the $\eta_\text{wind} = 0$ limit and instead use radial gas flows \citep[e.g.,][]{Lacey1985, Spitoni2011, Bilitewski2012} to adjust the metallicity gradient in their models.
Like ejection \citep[e.g.,][]{Weinberg2017,WellerWeinberg2026}, radial gas flows tend to lower the ISM metallicity and steepen the abundance gradient \citep[e.g.,][]{Portinari2000, Johnson2025solo}, leading to similar outcomes.
For our purposes, the choice of ejection over radial gas flows is purely practical.
Radial gas flows require assumptions regarding the physical processes driving the flow (see, e.g., the analytic models in \citealt{Johnson2025solo}), whereas our ejection rates can be simply asserted.

\subsection{The Primordial Helium Isotope Ratio}
\label{sec:impcosmo}

\begin{figure*}
\begin{center}
\includegraphics[width=1.5\columnwidth]{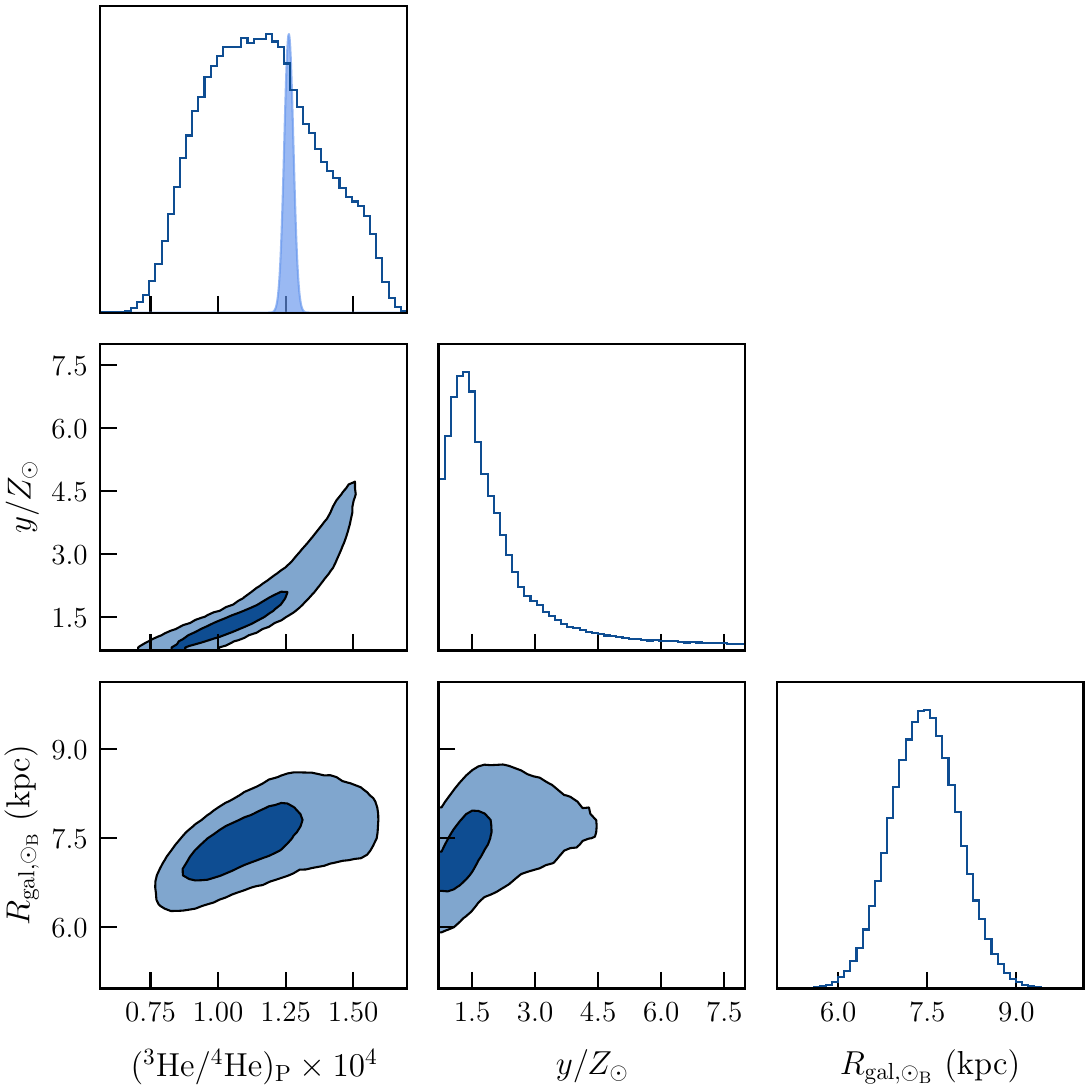}
\end{center}
\caption{The main diagonal elements show the 1D posterior distribution of each model parameter (histograms), while the remaining panels show the 68\% and 95\% confidence contours (dark and light shades, respectively) to illustrate the covariance between the model parameters. In the top left panel, the thin filled distribution represents the primordial $(^{3}{\rm He}/^{4}{\rm He})_{\rm P}$ value based on BBN calculations that assume the Standard Model of particle physics, the Standard cosmological model, and the baryon density of the Universe reported by the \citet{Planck2018}.
\label{fig:cornerplot}}
\end{figure*}

Our grid of \texttt{VICE} models highlights that the \heiso\ ratio varies with both time and galactocentric radius. Thus, there are two ways that we can pin down the \heiso\ chemical evolution of the Milky Way: (1) determine the present day \heiso\ at a range of different Galactocentric radii; or (2) measure the \heiso\ at different epochs of the Milky Way's chemical evolution. We can also incorporate both of these sensitivities using a joint analysis to infer the primordial helium isotope ratio and a scale factor for the stellar yields. Specifically, our new \heiso\ measures sample a range of Galactocentric distance ($6.5\le R_{\rm gal}/{\rm kpc}\le8.4$), while the protosolar \heiso\ ratio offers a snapshot of the \heiso\ chemical evolution of the Milky Way roughly 4.6 billion years ago. In what follows, we simultaneously use all of these measurements to pin down the \heiso\ chemical evolution of the Milky Way.

All of our \vice\ calculations are initialised with the same parameters as described in Section~\ref{sec:gce}. Each simulation is also initialised with a primordial helium isotope ratio and a value for the scale of the stellar yields.
We compute a grid of simulations that cover a wide range of this parameter space, including $0.5\times10^{-4}\le({\rm ^{3}He/^{4}He})_{\rm P}\le1.7\times10^{-4}$, in steps of $0.2\times10^{-4}$ and $0.7\le y/Z_{\odot} \le8.7$, in steps of $1$. Our implementation of radial migration means that stars will enrich the interstellar medium at an $R_{\rm gal}$ that is different from their birth $R_{\rm gal}$. One of the impacts of radial migration is that our models display small stochastic variations of \heiso\ as a function of $R_{\rm gal}$. To smooth out these effects, we perform 1000 random realisations of each model, whereby the radial migration of stars is randomised. We use these realisations to determine the average \heiso\ value and its intrinsic dispersion as a function of $R_{\rm gal}$.

We linearly interpolate our model grid and sample our parameter space using a Markov chain Monte Carlo (MCMC) procedure to find the most likely parameters that reproduce the available data. We also note that the aforementioned intrinsic model dispersion due to the effects of radial migration is included as a model uncertainty in our MCMC calculation. Our present-day absorption line measures of \heiso\ are compared to the final time snapshot of the simulations ($t=13.2~{\rm Gyr}$), which is chosen to be present age of the Milky Way. To model the protosolar \heiso\ ratio we use the $t=8.6~{\rm Gyr}$ \texttt{VICE} snapshot to be consistent with the expected age of the Solar System ($t_{\odot}=4.5682~{\rm Gyr}$; \citealt{BouvierWadhwa10}). We also note that the birth Galactocentric radius of the Sun is different from the present day Galactocentric radius, due to the combined effects of radial heating and angular momentum diffusion \citep{Minchev2018,Frankel2020}. In our MCMC calculation, we adopt a Gaussian prior on the birth Galactocentric radius of the Sun ($R_{\rm gal,\odot_{B}}$) with a mean value $\mu_{\odot_{B}}=7.3~{\rm kpc}$ and standard deviation $\sigma_{\odot_{B}}=0.6~{\rm kpc}$, which is based on the semi-empirical result by \citet{Minchev2018}. For our other model parameters, we adopt uniform priors on the primordial helium isotope ratio ($0.5\times10^{-4}$, $1.7\times10^{-4}$) and the scale of the stellar yields ($0.7$, $8$). We initialise the MCMC with 100 walkers each taking 50,000 steps, with values drawn from their prior distribution; for (\heiso)$_{\rm P}$ and $y/Z_{\odot}$ the walkers uniformly sample the prior distribution, while the birth Galactocentric radius of the Sun is initialised with a Gaussian random realisation of the prior distribution. To ensure convergence of the chains, we compute the integrated autocorrelation time. We adopt a burn-in of 1000 steps (far larger than the autocorrelation time) and thin the results by 15, which is about half the autocorrelation time.

The results of this calculation are shown in Fig.~\ref{fig:cornerplot}. The best-fit values and 68\%\ confidence intervals are:
\begin{eqnarray}
    (^{3}{\rm He}/^{4}{\rm He})_{\rm P}\times10^{4} =& 1.15^{+0.24}_{-0.21} \qquad {\rm (68\%)}\\
    y/Z_{\odot} =& 1.6^{+1.5}_{-0.6} \qquad {\rm (68\%)}
\end{eqnarray}
However, note that there is significant covariance between these two model parameters (see middle left panel of Fig.~\ref{fig:cornerplot}). Further information is needed to reduce this covariance, such as additional measurements in the more metal-poor outskirts of the Milky Way, combined with improved measures of the absorbers reported herein. The current confidence intervals quoted above are dominated by the statistical errors of the pre-solar value as well as the \tet\ and HD\,319718 measurements. Improving the \heiso\ measurement precision of the absorber towards HD\,319718 by a factor of 4 would result in a 14\%\ determination of the primordial helium isotope ratio. Such an improvement should also be matched with additional measurements at $R_{\rm gal}\gtrsim10~{\rm kpc}$, where the BBN contribution of \heiso\ is $\gtrsim75\%$. It would also be beneficial to map the Galactic distribution of non-metal isotopes, including \heiso, to pin down the parameters of the GCE models \citep{Johnson2025}.

Our measurement of the primordial helium isotope ratio is consistent with the Standard Model value, $(^{3}{\rm He}/^{4}{\rm He})_{\rm P} = (1.262\pm0.017)\times10^{-4}$, shown by the light blue filled region in the top left panel of Figure~\ref{fig:cornerplot}. We also note that our preferred value of the stellar yield scale is in good agreement with most stellar evolution and supernova models \citep[e.g.,][]{Woosley1995, Nomoto2013, Seitenzahl2013, Limongi2018}, which favour the range $1\lesssim y / Z_\odot \lesssim3$ (see discussion in, e.g., \citealt{Weinberg2024}).

\begin{figure*}
\includegraphics[width=\textwidth]{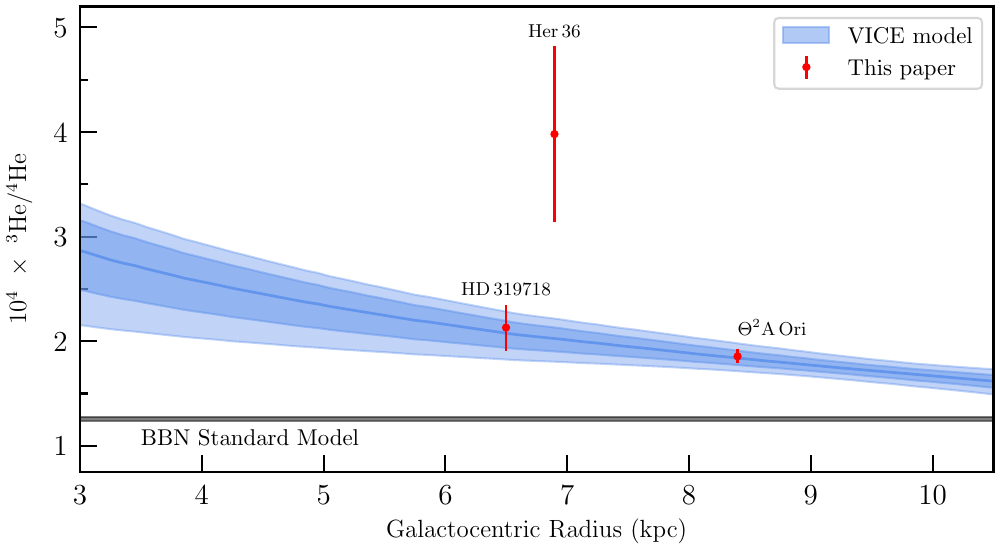}
\caption{Measurements of the \heiso\ isotope ratio (symbols with $1\sigma$ error bars) determined from our analysis (left to right, HD\,319718, Her~36, and \tet). Our best-fitting \vice\ model to the available measurements is shown by the central dark blue line, while the dark and light shaded bands represent the 68\% and 95\% confidence regions of the best fit model. The horizontal grey band shows the Standard Model value of the primordial helium isotope ratio, $(^{3}{\rm He}/^{4}{\rm He})_{\rm P}=(1.262\pm0.017)\times10^{-4}$.
\label{fig:rgal}}
\end{figure*}

We find that our fiducial model provides an accurate fit to the present day Milky Way measurements of \heiso\ (Fig.~\ref{fig:rgal}). We also find that the inferred protosolar value of this model (\heiso$=1.682\pm0.059$) reproduces the measured value (\heiso$=1.66\pm0.05$). Our model would be further informed by including a high precision measurement of the protosolar D/H abundance \citep{Johnson2025}, which is currently not available. Finally, we point out that our measurement of Her\,36 is above the inferred model by $2.3\sigma$. Given that the \HetIs\ absorption is blended with the right wing of the much stronger \HefIs\ absorption, it is partially affected by blending and the continuum determination. Further data of ${\rm S/N}>1500$ (roughly three times higher than the current data) are required to resolve the blending between the blue edge of \HetIs\ and the red edge of \HefIs. If blending and a misplaced continuum cannot explain the elevated \heiso\ value towards Her\,36, we may need to explore astrophysical explanations, such as the impact of radiative transfer effects on the optically thick \HefIs\ line profile, real intrinsic variations of the \het\ production in the Milky Way as a function of Galactocentric distance, or departures from our assumption that charge transfer reactions ensure $N(^{3}$\HeIs)/$N(^{4}$\HeIs) closely follows the intrinsic helium isotope ratio, \heiso~=~$N(^{3}$\HeIs)/$N(^{4}$\HeIs).

\subsection{Yield Scale of the Milky Way}

Until future measurements of D/H or \heiso\ become available for the Milky Way, it may be difficult to pin down the primordial \heiso\ ratio and the scale of the stellar yields. 
Uncertainties in stellar evolution models make it challenging to pin down the overall scale of metal yields (see discussion in Section \ref{sec:intro}), with \het\ production and its associated uncertainties being particularly relevant to the present paper.
\het\ yields are sensitive to prescriptions for convection and extra mixing in stellar interiors, and recent observations of planetary nebulae challenge expectations based on ``standard'' assumptions (see discussion in, e.g., \citealt{BanBal21} and \citealt{Balser2022}).
In the meantime, to further understand the scale of the helium isotopes stellar yield, we place a Gaussian prior on the Standard Model primordial abundance, (\heiso)$_{\rm P}=(1.262\pm0.017)\times10^{-4}$, and repeat our MCMC calculation to infer the yield scale. Adopting the same set of MCMC parameters as above, we infer $y/Z_{\odot}=2.12^{+0.31}_{-0.29}$ (68\% confidence).
Our statistical inferences of $y / Z_\odot$ should reflect information encoded in the \heiso\ ratios in the ISM \textit{only}, since our models are calibrated to asymptotically approach the present-day O/H in the ISM (see discussion in Section \ref{sec:gce}).
\par
\citet{Weinberg2024} computed a lower value of $y/Z_\odot = 1$ based on the mean Fe yield of Type II supernovae measured by \citet{Rodriguez2023}. Although this is somewhat lower than the value we infer, there are several possibilities that could bring these measurements into closer agreement. For example, our models assume that accreting gas is zero metallicity.
Relaxing this assumption and allowing non-pristine accretion would lead us to infer a lower yield scale.
Without drawing quantitative conclusions, \citet{Johnson2025RMG} demonstrated that GCE models using $y/Z_\odot \gtrsim 2$ more readily reproduce the surprisingly minimal decline in metallicity toward old stellar ages \citep[e.g.,][]{Willett2023, Gallart2024, Roberts2025}.
However, they also showed that metal-rich accretion should play a role in flattening the age-metallicity relation in stars, making it challenging to favour a given value of $y / Z_\odot$ based on their comparisons. \citet{WellerWeinberg2026} highlight the importance of obtaining new measurements of \heiso\ at somewhat higher metallicities (lower $R_{\rm gal}$) to distinguish between the relative importance of outflows and the stellar yield scale.

\section{Conclusions}\label{sec:conc}
We report the discovery of two gas clouds that exhibit metastable helium absorption, based on archival optical observations. Given the high \HeIs\ column density and quiescent kinematics of the absorbing clouds, we obtained new observations with the CRIRES$^{+}$ spectrograph to resolve the isotopic absorption lines of metastable helium, and obtain the first measures of \heiso\ of these clouds. To complement these data, we also obtained new observations of a previously known \HetIs\ absorber. The main conclusions of this work are as follows:

\smallskip

(i) Using two time-separated measurements ($\Delta t\sim8~{}\rm years$) of weak \HeIs\ absorption towards \tet, we find that the absorption does not vary with time to within the measurement error, and report a $2\sigma$ upper limit ${\rm d}\log_{10}\,[N({\rm He\,\textsc{i}}^{*})/{\rm cm}^{-2}]/{\rm d}t\leq7.2\times10^{-4}~{\rm dex~yr}^{-1}$ (equivalent to $<0.17\%~{\rm yr}^{-1}$ when expressed as a linear quantity). We also place a tight limit on the possible variation of the Doppler parameter along this sightline, ${\rm d}b/{\rm d}t \leq 11~{\rm m~s^{-1}~yr^{-1}}$ ($2\sigma$ confidence). This can also be expressed as a limit on the variation of the kinetic temperature of the absorbing medium, ${\rm d}T/{\rm d}t \leq 36~{\rm K~yr^{-1}}$.

\smallskip

(ii) We report the discovery of metastable helium absorption towards HD\,319718 (in the open cluster Pismis 24) and Her\,36 (in the Lagoon Nebula). We find that the \HeIs\ absorption is present in multiple components, and these individual components are well-described by Voigt profiles, provided the absorption line is optically thin. Optically thick absorption lines exhibit mildly asymmetric line profiles, and are slightly blueshifted ($\lesssim2~{\rm km~s}^{-1}$) relative to the optically thin absorption lines. Since the measurement of \heiso\ is based on weak lines, the minor blueshift of the optically thick near-infrared \HefIs\ line relative to the optically thin \HefIs\ lines at optical wavelengths does not affect the measured \heiso.

\smallskip

(iii) We also obtained new, deep observations of \tet, and have improved the \heiso\ measurement precision of this sightline by a factor of $\sim2$.

\smallskip

(iv) Using the recent Milky Way galactic chemical evolution models described by \citet{Johnson2025RMG}, we computed a grid of \vice\ calculations that cover a wide range of primordial \heiso\ values, and a range of scale factors for the stellar yields. We use an MCMC procedure to find the model parameters that best reproduce our observations. Based on our data and modelling, we estimate a primordial helium isotope ratio, (\heiso)$_{\rm P}=(1.15^{+0.24}_{-0.21})\times10^{-4}$ ($\sim20$ percent precision), which agrees with the value derived from BBN calculations that assume the Standard Model of particle physics in combination with the baryon density inferred from the cosmic microwave background temperature fluctuations.

\smallskip

(v) To learn more about the post-BBN production of \het, we place a prior on the primordial helium isotope ratio (based on the Standard Model in combination with \obhh\ from the CMB) to infer the scale of the stellar yields, with a 68\% confidence interval of $y/Z_{\odot}=2.12^{+0.31}_{-0.29}$. This value is somewhat higher than other probes of the yield scale, but could be reduced by modifying the chemical composition of the accreted pristine gas to include metals and a post-BBN contribution of \het\ and \hef. New measurements of \heiso\ and D/H across a wide range of galactocentric radii in the Milky Way will break the degeneracy between (\heiso)$_{\rm P}$ and $y/Z_{\odot}$. This degeneracy could also be broken with additional measurements of \heiso\ in more metal-poor environments, such as the Large and Small Magellanic Clouds.

\smallskip

Ultimately, the \heiso\ ratio provides us with a rare probe of cosmology during the first day after the Big Bang. Even though our measurements are currently limited to the Milky Way, we have shown here that this technique has the potential to deliver high precision measures of \heiso\ ($\lesssim4\%$). We find that the Milky Way \heiso\ ratio is elevated by just $\sim40\%$ above the primordial value expected for the Standard Model in combination with the \obhh\ derived from the CMB. We propose that future measures of \heiso\ should focus on pushing this work towards more metal-poor environments, including nearby metal-poor star-forming galaxies, and transient objects such as supernovae and gamma-ray bursts that explode in metal-poor environments (e.g. \citealt{Fynbo14}). It is also important to obtain complementary measures of \heiso\ along new sightlines that cover a range of Galactocentric radius in the Milky Way. This will allow us to understand the post-BBN production of \het, the galactic chemical evolution of the Milky Way, and potentially infer the primordial \heiso\ from the Milky Way's more metal-poor outskirts. The forthcoming generation of extremely large telescopes, equipped with high resolution spectrographs, are ideally suited to address these goals (e.g. ELT/ANDES; \citealt{Martins2024}).

\begin{acknowledgments}
We thank an anonymous referee for their timely review of our manuscript, and for the many helpful comments offered that allowed us to clarify various aspects of our analysis.
The authors would like to thank David Weinberg for providing insightful feedback that improved the discussion of our results in the context of the Milky Way's chemical evolution. This paper is based on observations collected at the European Organisation for Astronomical Research in the Southern Hemisphere, Chile (VLT program IDs: 081.D-0854(B), 091.C-0851(A), 107.22U1.001, 110.23QS.001, 110.23QS.002, 112.25D8.001, 112.25D8.002, 194.C-0833(A), 194.C-0833(D)). We are most grateful to the staff astronomers at the VLT for their assistance with the observations.
During this work, RJC was supported by a
Royal Society University Research Fellowship.
RJC acknowledges support from STFC (ST/T000244/1, ST/X001075/1).
JWJ acknowledges financial support from a Carnegie Theoretical Astrophysics Center postdoctoral fellowship.
This research has made use of NASA's Astrophysics Data System.
\end{acknowledgments}

\vspace{5mm}
\facilities{VLT(CRIRES and UVES)}

\software{ALIS \citep{Cooke2014},
          astropy \citep{Astropy2013,Astropy2018},
          corner \citep{corner},
          emcee \citep{emcee},
          matplotlib \citep{matplotlib},
          numpy \citep{numpy},
          PypeIt \citep{PypeIt},
          scipy \citep{scipy},
          sympy \citep{SymPy},
          VICE \citep{Johnson2021}.
          }

\bibliography{mnbib}{}
\bibliographystyle{aasjournal}

\end{document}